\documentclass[a4paper,fleqn]{cas-dc}

\usepackage[numbers]{natbib}
\usepackage{amsmath}

\def\tsc#1{\csdef{#1}{\textsc{\lowercase{#1}}\xspace}}
\tsc{WGM}
\tsc{QE}
\tsc{EP}
\tsc{PMS}
\tsc{BEC}
\tsc{DE}
\begin{document}

%\let\WriteBookmarks\relax
%\def\floatpagepagefraction{1}
%\def\textpagefraction{.001}

\shorttitle{A compartmental model for cyber-epidemics}

\shortauthors{D. Aleja, G. Contreras-Aso et~al.}

\title[mode = title]{A compartmental model for cyber-epidemics} 
%\tnotemark[]

%\tnotetext[1]{This document is the results of the research project funded by the National Science Foundation.}

\author[1,2,3]{ D. Aleja} 
\cormark[1]
\author[1,2]{ G. Contreras-Aso}
\cormark[1]
\author[1,2]{ K. Alfaro-Bittner}
\author[1,2]{ E. Primo}
\cormark[2]
\ead{eva.primo@urjc.es}
\author[1,2,3]{ R. Criado}
\author[1,2,3]{ M. Romance}
\author[1,2,4,5]{ S. Boccaletti}
%\cormark[2]
%\fnmark[1,3]
%\ead{stefano.boccaletti@gmail.com}

\address[1]{Universidad Rey Juan Carlos, Calle Tulip\'an s/n, M\'ostoles, 28933, Madrid,Spain}

\address[2]{Laboratory of Mathematical Computation on Complex Networks and their Applications (LaCoMaRCA), Universidad Rey Juan Carlos, Calle Tulip\'an s/n, M\'ostoles, 28933 , Madrid, Spain}

\address[3]{Data, Complex networks and Cybersecurity Research Institute, Universidad Rey Juan Carlos, Plaza Manuel Becerra 14, 28028, Madrid, Spain}

\address[4]{Moscow Institute of Physics and Technology, Dolgoprudny, 141701, Moscow, Russian Federation}

\address[5]{CNR - Institute of Complex Systems, Via Madonna del Piano 10, I-50019, Sesto Fiorentino, Italy}

\cortext[cor2]{These Authors equally contributed to the Manuscript}
\cortext[cor1]{Corresponding author}
%\fntext[fn1]{These Authors equally contributed to the Manuscript}

\begin{abstract}
In our more and more interconnected world, a specific risk is that of a cyber-epidemic (or cyber-pandemic), produced either accidentally or intentionally,
where a cyber virus propagates from device to device up to undermining the global Internet system with
devastating consequences in terms of economic costs and societal harms
related to the shutdown of essential services.
We introduce a compartmental model for studying the spreading of a malware and of the awareness of its incidence
through different waves which are evolving on top of the same graph structure
(the global network of connected devices). This is realized by considering vectorial compartments made of two components,
the first being descriptive of the state of the device with respect to the new
malware's propagation, and the second accounting for the awareness
of the device's user about the presence of the cyber threat. By introducing suitable transition rates between such  compartments,
one can then follow the evolution of a cyber-epidemic from the moment at which a new virus is seeded
in the network, up to when a given user realizes that his/her device has suffered a damage and consequently starts a wave of
awareness which eventually ends up with the development of a proper antivirus software.
We then compare the overall damage that a malware is able to produce in Erd\H{o}s-R\'enyi and scale-free network architectures
for both the case in which the virus is causing a fixed damage on each device and the case where, instead, the virus is
engineered to mutate while replicating from device to device.
Our result constitute actually the attempt to build a specific compartmental model whose variables and parameters
are entirely customized for describing cyber-epidemics.
\end{abstract}

%Research highlights
%\item A compartmental model for cyber-epidemics
%\item Epidemic spreading on networks
%\item Network resilience to virus propagation
%\item Effects on network topologies for malware spreading

% Keywords

\let\WriteBookmarks\relax
\def\floatpagepagefraction{1}
\def\textpagefraction{.001}

\maketitle

The study of compartmental models started already at the beginning of the 20th century~\cite{ross1,ross2,ross3,kermack},
and soon became a subject of great, recent, interest that has attracted the attention of many epidemiologists.
In analogy with cellular automata~\cite{wolfram}, these models
consider a networked population of individuals, each one of them described by a state whose discrete values
are labeled by {\it compartments}. Individuals may then progress between compartments through given transition probabilities,
which allow the time-discrete evolution of the population during, for instance, the spreading
of infectious diseases and/or rumors and social contagion.
Physicists became interested in these models when it was pointed out that epidemiological
processes can be regarded as percolation like processes~\cite{grassberger}.

Starting from the seminal work by Pastor–Satorras and Vespignani~\cite{pastor}, the last twenty years have seen a
burst of activity on understanding the effects of a network topology on the rate and patterns of the disease spread.
A lot of studies have tried indeed to predict things such as the total number of infected individuals, or the duration of an epidemic,
and to estimate various relevant parameters such as the reproductive number.
Moreover, and especially in relation to the recent COVID19 world pandemic crisis, these models
have been used to assess the effects of public health interventions and/or to quantify the efficiency of issuing
a limited number of vaccines in a given population~\cite{noiCSF1,noiCSF2}.

In particular, the SIR model~\cite{murrayetal1,murrayetal2} is one of the simplest compartmental models, and describes the evolution of diseases resulting
in the immunization or death of the infected individuals. The model assumes that, at each time, each individual can be in one of three possible
compartments: susceptible (denoted by $S$), infected ($I$), or removed ($R$). The susceptible units of the network are those healthy persons that can
develop the disease if they get in contact with infected individuals. Once an individual contracts the infection, it moves into the infected
(and infective) compartment, and then, after some time, into the removed compartment, which indicates that the individual
cannot catch the disease anymore (or passes it on), due to a lasting resistance conferred by the recovery (or because it dies).

\begin{figure}
    \centering
    \includegraphics[]{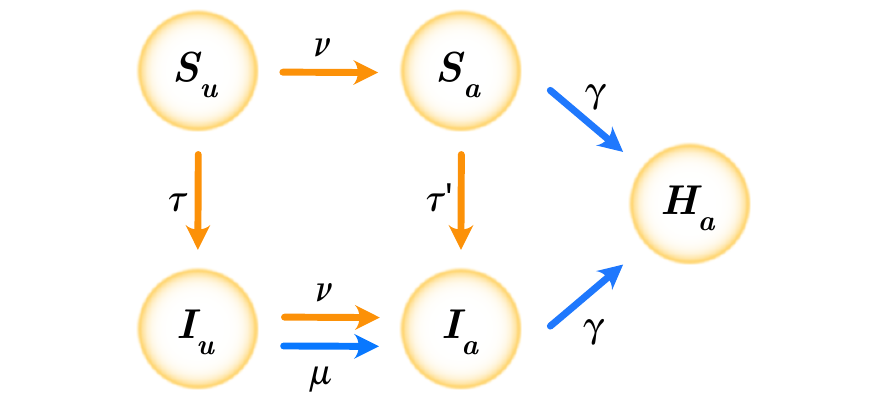}
    \caption{{\bf  The compartmental model for cyber-epidemics.} Units of the network can be in one of four possible states before recovery: susceptible-unaware ($S_u$), susceptible-aware ($S_a$), infected-unaware ($I_u$), and infected aware ($I_a$). The recovery or healed status ($H_a$) is by  nodes that, after getting aware of the existence of the malware, applies a suitable operation on the system. Orange (blue) arrows stand for contact-based (localized spontaneous) transitions.% Transition rates used in our simulations are $\tau = 0.0055$, $\tau' = \tau/10$, $\nu = 0.011$, $\gamma = 0.03$, and $\mu = x\cdot H(x)$, where $x=\nu (d - \theta)$ and $H(x)$ is the Heavyside theta function, whose purpose preventing transitions for viruses whose current damage $d$ is below a given threshold $\theta$.
    }
    \label{compartmental_model}
\end{figure}

Just as a viral pandemic proliferates inside a population of individuals, a ransomware (or other malware) software can spread
within the global Internet. As technology is today globally interconnected, a cyber virus can indeed propagate from device to device, with devastating consequences.
Cyber attacks on infrastructure services are currently on the rise, and hackers are exploiting the use of the Internet of Things which creates millions of new vulnerability points in all critical infrastructures. A specific risk is therefore the occurrence of a cyber-epidemics (or cyber-pandemic), produced either accidentally or intentionally, and undermining the global Internet system up to the need of its lockdown. The World Economic Forum~\cite{worldeconomic} predicted that a single day without Internet would cost around 50 billion USD globally, without even considering the societal harm related to the shutdown of essential services.

%\begin{table*}[h!]
%    \centering
%    \begin{tabular}{llll}
%    \hline
%        Symbol & Description & Range & Chosen value \\
%        \hline
%        $\tau$ & Infection rate & [0,1] & 0.0055 \\
%        $\nu$ & (Contact-based) awareness parameter & [0,1] & 0.011 \\
%        $\mu_0$ & (Spontaneous or local) awareness parameter & [0,1] & 0.011 \\
%        $\gamma$ & Recovery parameter & [0,1] & 0.03 \\
%        $\rho_0$ & Fraction of population initially infected & [0,1] & 0.01 \\
%        \hline
%    \end{tabular}
%    \caption{Summary of the parameters involved in the model.}
%    \label{parameters}
%\end{table*}

Since the first attempt to describe the spreading of computer viruses \cite{firstvirus}, several other studies have
tried to adapt classical, global or networked, compartmental models  with the aim not only of investigating
the propagation cycle of cyber viruses but also to evaluate the effectiveness of possible security countermeasures~\cite{modelcyber1,modelcyber2,modelcyber3,modelcyber4,modelcyber5,modelcyber6,modelcyber7}.
There is, however, a fundamental difference between the spreading of a biological virus within a population and a cyber-epidemic.
In the former case, indeed, each individual is passive actor of the game, contracting the disease and recovering from it due to the action
of its immune system. The latter case can instead be seen as the struggle between two kind of actors:
the ones who intentionally program the malicious code and try to seed it within the global Internet (that we will call
from here after as the ``bad team'') and those (that we will call
from here after as the ``good team'')
who are instead engaged in programming the corresponding antivirus code after
becoming aware of the presence of the new malware in the network, and in spreading it to all network's users.

In this paper, we introduce a novel compartmental model, whose variables and parameters are entirely customized for the case of
a cyber-epidemic. We will then compare the spreading of a malware on top of Erd\H{o}s-R\'enyi (connected) and scale-free network architectures
for both the case in which the virus is producing a fixed damage on each device and the case where, instead, the virus is
engineered to mutate while replicating from device to device.

In our model, both the spreading of malware and awareness occurs through waves developing and evolving within the same graph structure
(the global network of connected devices). Notice that this differentiates our approach from that of Ref.~\cite{multilayerarenas}
which describes  the dynamical interplay of a virus and of an awareness level on top of multiplex networks.
In our case, instead, we consider two dimensional vectorial compartments, the first component of which being the state of the device with respect to the new
malware's propagation ($S$ for susceptible, $I$ for infected, and $H$ for healed) while the second component accounts for the awareness
of the device's user about the presence of the new virus (``a'' for aware and ``u'' for unaware). Transition rates between such vectorial compartments
are then defined to properly model the behavior of a cyber-epidemic from  the moment at which a new (i.e., yet unknown to the good team) virus is seeded
in the network by the bad team, up to the moment at which a given user realizes that his/her device has suffered a damage and consequently starts a wave of
awareness which eventually ends up with reaching one of the units of the good team. Then, the good team develops a proper antivirus software that the
aware units download from the net for healing their devices.

In other words, the framework we adopt is somewhat like a vectorial version of the SIR model, where each susceptible or infected node is also either aware or unaware of the existence of the virus. This means that one has four possible states before recovery: susceptible-unaware ($S_u$), susceptible-aware ($S_a$), infected-unaware ($I_u$), and infected aware ($I_a$), see Fig.~\ref{compartmental_model}. Finally, the recovery or healed status ($H_a$) can be reached only by a node (a device) whose user, after getting aware of the existence of the virus, applies a suitable operation on the system, for instance installing an antivirus.

The transition probabilities between such states are mediated by four fundamental parameters, whose meaning is directly linked to specific processes occurring during a
cyber epidemic:

\begin{itemize}
    \item $\tau$ is the equivalent of the standard infection parameter in the SIR model. Here, it accounts for the contact-based transition rate from the susceptible to the  infected status. In the case of $S_a$ individuals, a different infection rate is used ($\tau'=\tau/10$), due to the fact that awareness  makes individuals reluctant to be in contact with infected sites/computers (for instance, email phishing campaigns normally raise alarms in media, making people more concerned about clicking on dubious links);

    \item $\nu$ is the rate of spreading of awareness due to network contacts, i.e., the parameter who rules the contact-based transition from the unaware to the aware status;

    \item $\mu$ accounts for the individual awareness parameter, i.e., the rate at which an infected unaware user actually notices that his/her device has suffered a damage (larger than a given threshold) due to having been infected by a new cyber virus, and consequently turns to the infected aware state and simultaneously starts signaling the existence of a new  virus circulating in the net;

    \item $\gamma$ is the recovery/healing parameter, i.e., the rate at which an aware (either susceptible or infected) individual gets healed by purging of the system via the installation of a proper antivirus.
\end{itemize}
\begin{table*}[h!]
    \centering
    \begin{tabular}{llll}
    \hline
        Symbol & Description & Range & Chosen value \\
        \hline
        $\tau$ & Infection rate & [0,1] & 0.0055 \\
        $\nu$ & (Contact-based) awareness parameter & [0,1] & 0.011 \\
        $\mu_0$ & (Spontaneous or local) awareness parameter & [0,1] & 0.011 \\
        $\gamma$ & Recovery parameter & [0,1] & 0.03 \\
        $\rho_0$ & Fraction of population initially infected & [0,1] & 0.01 \\
        \hline
    \end{tabular}
    \caption{Summary of the parameters involved in the model.}
    \label{parameters}
\end{table*}
Two types of transitions may therefore occur in our model: contact-based transitions and individual (or contact-indepen\-dent) ones. In individual transitions, the change of state is fully independent of the states of the rest of the devices, but is only due to the perception of the damage caused to the device (in the case of the passage from the state $I_u$ to the state $I_a$ at rate $\mu$) and/or the user's interest in using an antivirus (in the case of the passage from any aware state to the state $H_a$ at rate $\gamma$).

\begin{figure}
    \centering
    \includegraphics[width=0.99\linewidth]{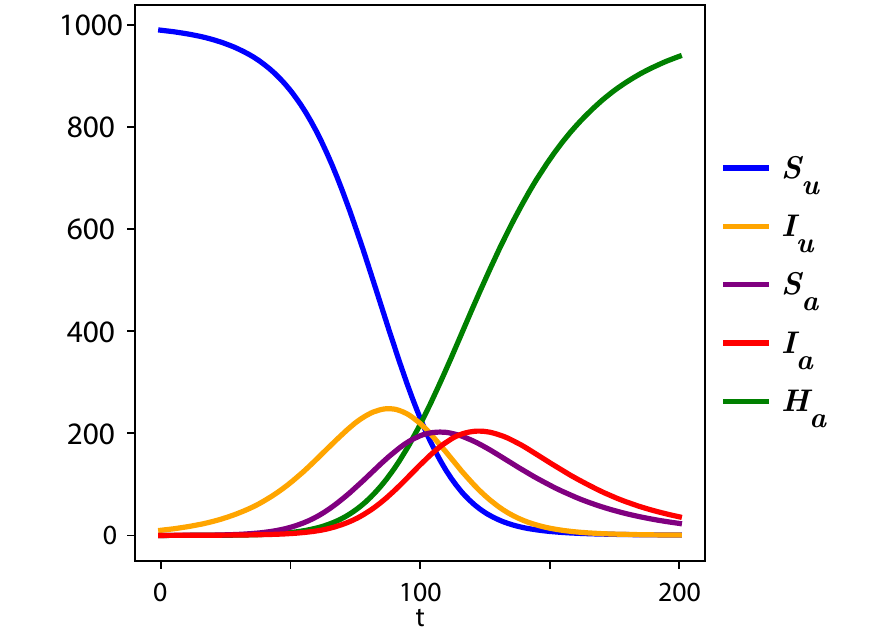}
    \caption{{\bf  The typical cycle of a cyber-epidemic.} The total number of $S_u$, $S_a$, $I_u$, $I_a$ and $H_a$ units (see legend for color code) vs. time, during a typical cyclic evolution of the virus. Simulations refer to an ensemble average over 1000 realization of an Erd\H{o}s-R\'enyi random network with $N=1000$ nodes, average degree $\langle k\rangle=10$, threshold $\theta=0.2 $ and constant damage $d=0.3$. See Table~\ref{parameters} for the values of the other parameters used. It is possible to distinguish three different phases of the cycle: an initial phase of no awareness where more and more nodes pass from the $S_u$ to the $I_u$ state; a second stage, the awareness phase starting at around $t \sim 30$, when a given $I_u$ node becomes $I_a$ and ignites an awareness wave so that the total number of both $I_a$ and $S_a$ grows by contact transitions; a final healing stage where all nodes turn to the $H_a$ state.}
    \label{fig_cycle}
\end{figure}

In our simulations, we initially prepare the network with all its units in the $S_u$ state. Then, a new malware is seeded by the bad team, i.e., one node (or a small group of nodes) is turned to the state $I_u$, and contact propagations start, yielding an initial spreading of the virus. At a second time, i.e., when one of the infected users gets aware of the damage produced by the cyber virus in its device, a second wave (which spreads awareness in the same network) starts due to an initial local transition from the state $I_u$ to the state $I_a$. Eventually, the entire cycle of the cyber-epidemic takes place, with an end state of the network where all its units are in the $H_a$ state \cite{specifications}.

In order to properly monitor how damaging a cyber-epidemic can be in a network of devices, we introduce as a further parameter the damage $d\in [0,1]$ caused when a device is infected by the virus, and we quantify the total damage caused to the system as the sum of the damages made to each single device.  Finally, as the $\mu$ parameter is totally related to the damage produced in a device, it is appropriate to define it to be proportional to the $d$ parameter. In addition, in order to account for the fact that there may be situations where the virus damage is not perceived at all by the user and the device continues to function correctly even with high damage or, on the opposite, cases (like, for instance, in high sensitivity infrastructures) where just a little damage constitutes an alarm on the existence of the virus, the parameter $\mu$ is activated depending on whether it exceeds a threshold $\theta\in [0,1]$. The latter lead to the following expression with $\mu_0 \in[0,1]$:
\begin{equation*}
\mu=\begin{cases}
\displaystyle \mu_0(d-\theta) \;\; \hbox{if} \;\; d\geq \theta, \\
\displaystyle \mu=0  \;\; \hbox{if}\;\;  d< \theta.
\end{cases}
\end{equation*}

As for the threshold $\theta$, the higher it is the more damage individual systems can withstand before realizing the presence of the virus. If the threshold is at a low value (as, for instance, in high security infrastructures or clusters) a damage as small as the deletion of a couple of files in the system is already sufficient to trigger an alarm. If, instead, the threshold is high, then a few files missing would be taken as an error, but a large system failure would actually be understood as the consequence of a virus. Lower thresholds are therefore always preferable for protecting an overall system from damage, but in practice they are overcostly as they need more investments in infrastructure protection and even more employees to maintain it.

Fig.~\ref{fig_cycle} reports the total number of $S_u$, $S_a$, $I_u$, $I_a$ and $H_a$ units vs. time during a typical cyclic evolution of the virus, for an Erd\H{o}s-R\'enyi random network of $N=1000$ nodes, with average degree $\langle k\rangle=10$, $\theta=0.2$, $\tau=0.0055$, $\nu=0.011$, $\mu_0=0.011$, $\gamma=0.03$ and $\rho_0=0.01$ (being $\rho_0$ the fraction of population initially infected, which therefore means that the virus is initially seeded in 10 devices). From the figure, one can clearly distinguish the three different stages of the cycle: an initial stage of no awareness where more and more nodes are infected (passing from the $S_u$ to the $I_u$ state), the beginning of the awareness phase (around $t \sim 30$) when a given $I_u$ node becomes $I_a$ and starts an awareness wave so that the total number of both $I_a$ and $S_a$ grows by contact transitions, and the final healing stage where all nodes turn to the $H_a$ state.

In order to quantify the total injury produced in the system by the cyber virus during its cycle of evolution, we introduce the quantity $D/N$ accounting for the normalized sum of the individual damages suffered by each node. Its value is obtained with multiplying $d$ by the total number of infected nodes during a cycle (regardless on whether they are in the state $I_u$ or $I_a$), and dividing by $N$.
When $d$ is below the threshold $\theta$  the presence of the cyber threat is never detected and, as a consequence, the virus will propagate to all nodes in the network yielding $D/N = (N\cdot d)/N = d$. A non trivial behavior is instead observed for all values $d > \theta$, where the awareness mechanism is activated at a given time in the cycle.

As a first step, we face the problem of assessing how different topologies of the network react to a virus (causing a constant damage $d$ when infecting a device) at different values of the threshold  $\theta$.

\begin{figure*}
    \centering
    \includegraphics[width=0.99\linewidth]{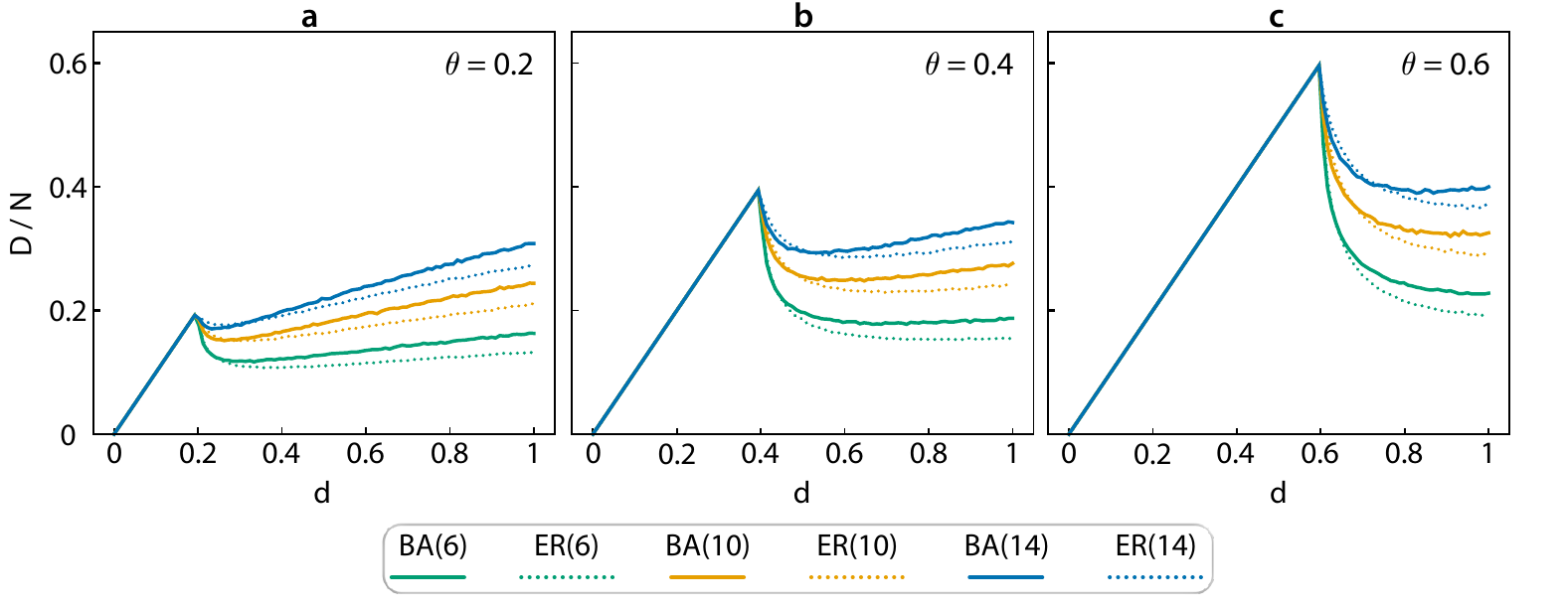}
    \caption{{\bf  Effects of network topology.} $D/N$ vs. $d$, for different choices of networks (Barab\'asi-Albert or Erd\H{o}s-R\'enyi) of mean degree 6, 10 and 14 (see legend at the bottom of the figure for color code) and different values of the threshold $\theta$: a) $\theta=0.2$, b) $\theta=0.4$, and c) $\theta=0.6$.}
    \label{fixed_damage}
\end{figure*}

With the aim of comparing homogeneous and heterogeneous topologies Fig.~\ref{fixed_damage} illustrates the results of our simulations, and reports $D/N$ vs. $d$ for Barab\'asi-Albert~\cite{bara1,bara2} scale free and Erd\H{o}s-R\'enyi~\cite{erdos} random networks, at different values of the mean degree $\langle k\rangle$ and different values of $\theta$.
A first, even though rather trivial, evidence is that the global damage is higher for higher values of $\theta$, indicating that the harder it is for nodes to become aware, the more free is the cyber threat to propagate without countermeasures.
Moreover, a noticeable difference is observed regarding the position of the maximum damage that a virus can deal in a given system, being it located at $d=1$ for low thresholds and at $d=\theta$ for high thresholds.
Finally, and more remarkably, the results shown in Fig.~\ref{fixed_damage} allow to conclude that Barab\'asi-Albert scale free (i.e., heterogenous) networks are more fragile, and more sensitive to the spreading of viruses causing a fixed damage than Erd\H{o}s-R\'enyi (i.e., homogenous) ones, so that a first conclusion is that engineering a network in a scale free topology renders it more vulnerable to cyber-epidemics, in line with what was already known about their structural fragility against intentional attacks~\cite{naturebara}.
On the other hand, it is seen that, regardless on the specific network topology, the higher is the average degree the higher is the graph's vulnerability.
Already at this qualitative level, a first conclusion can be drawn: if the threshold is high, inflicting maximal injury to a system requires some tuning on the side of the bad team (the virus designers), with the ideal value of the damage to be caused at each single device being that of the threshold. On the opposite, if the threshold is low, then a high base damage virus is able to inflict a huge damage on the network regardless on the specific threshold value.

\begin{figure}
    \centering
    \includegraphics[]{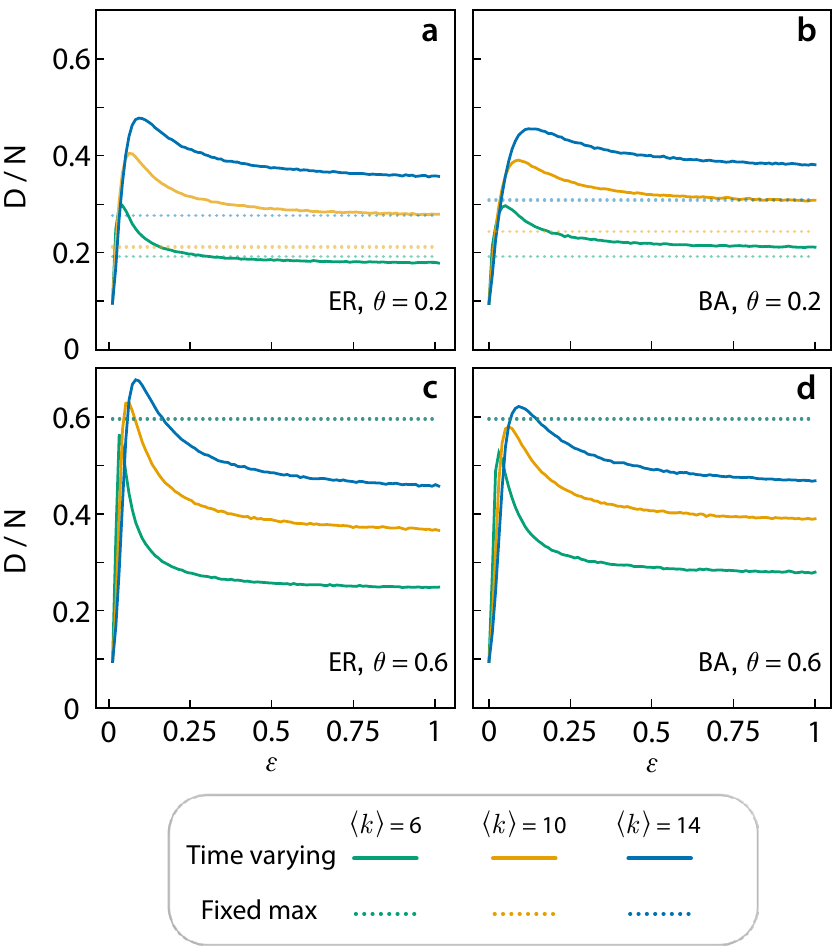}
    \caption{{\bf  Effects of time varying damage.} $D/N$ vs. $\varepsilon$ (see text for definition), for different choices of networks (Erd\H{o}s-R\'enyi in the left column and Barab\'asi-Albert in the right column) of mean degree 6, 10 and 14 (see legend at the bottom of the figure for color code) and different values of the threshold $\theta$: a,b) $\theta=0.2$, c,d) $\theta=0.6$. In all panels the maximal injuries achieved by a fixed strength virus (the global maximum of the curves of Fig.~\ref{fixed_damage}) are reported with horizontal dashed lines.}
    \label{fixed_vs_time_varying}
\end{figure}

We then move to consider the effects of a damage which is variable in time, that is, $d=d(t)$, where $t$ is a discrete time measuring the number of iterations in the model (a time unit being the lapse from one to another iteration of the networks' nodes). Namely, starting at $d_0=0.1$, the damage is increased in times as
\begin{equation}
    d(t)=\frac{d_0e^{\varepsilon t}}{1+d_0(e^{\varepsilon t} -1)},
    \label{equazcrescita}
\end{equation}
for some $\varepsilon>0$. Note that the function $d(t)$ is the logistic function with growth rate $\varepsilon$, initial population $d_0$ and carrying capacity $1$, so
that it displays an exponential behavior ($d(t)\approx d_0 e^{\varepsilon t}$) when $t\approx 0$ and approaches $1$ when $t\to \infty$. Choosing the logistic function as the damage function is ideal in the model, as the intentions of the bad team is precisely that of spawning as much damage as possible before the virus is detected by users.

Fig.~\ref{fixed_vs_time_varying} reports $D/N$ vs. $\varepsilon$, for both Erd\H{o}s-R\'enyi (left column) and Barab\'asi-Albert (right column) at different mean degrees and different values of $\theta$. For comparison, in the same figure the maximum possible injury caused to the system by a fixed strength virus is reported by horizontal dashed lines. One can easily see that, as a function of $\varepsilon$, $D/N$ displays an initial monotonic growth and an asymptotic behavior for $\varepsilon \to 1$, featuring a local maximum for some $0 < \bar \varepsilon <1$.
Remarkably, one can notice that for low values of the threshold, there is a range of $ \varepsilon $ for which the amount of injury inflicted to the system is actually higher than the maximum at constant base damage. This implies that the bad team has the option of causing a higher injury on a high security system by just engineering the virus with an internal clock, which would progressively increase the damage inflicted to infected devices. On the contrary, at high values of $\theta$ (i.e., when security is not so demanding), viruses whose base damage is given by Eq.~(\ref{equazcrescita}) cause an overall injury which is comparable with the maximal value at constant damage strength.

Finally, we briefly describe the scenario in which the virus is engineered to increase its base damage not due to an internal clock, but due to a mutation that occurs all the times the virus is transmitted from an infected device to a susceptible one. This is tantamount to say that, when a device is infected at time $t$, the damage caused depends on the story of the specific strain of the infecting virus. In other words, Eq.~(\ref{equazcrescita}) is substituted by
\begin{equation*}
    d(t_i)=\frac{d_0e^{\varepsilon t_i}}{1+d_0(e^{\varepsilon t_i} -1)} \;\; \hbox{with} \;\; d_0=0.1,
    \label{equazmutazione}
\end{equation*}
where the discrete variable $t_i$ is now the number of times that the infective strain propagated before reaching the actual device.
Therefore, the damage caused is now explicitly a function of the specific history of the strain propagation. Moreover, for the $I_u \to I_a$ local transition, the parameter $\mu$ is determined by looking at the base damage received by the device at the moment of its contagion.

\begin{figure}
    \centering
    \includegraphics[]{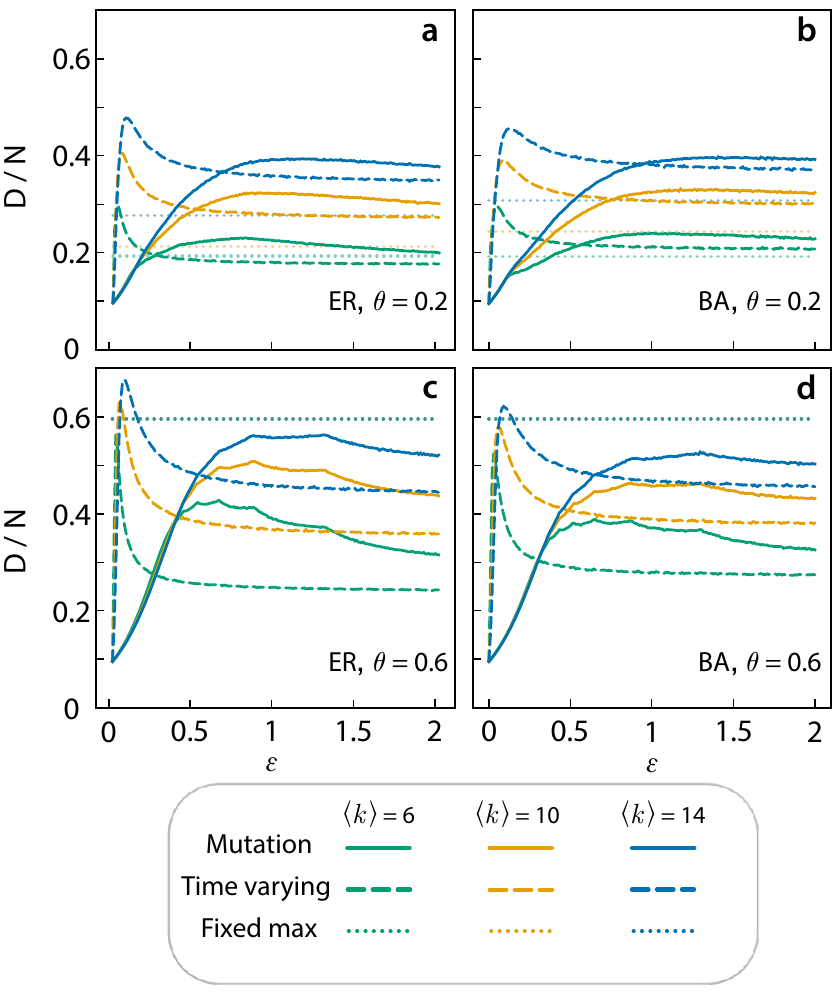}
    \caption{{\bf  Mutating virus.} $D/N$ vs. $\varepsilon$ (see text for definition), for different choices of networks (Erd\H{o}s-R\'enyi in the left column and Barab\'asi-Albert in the right column) of mean degree 6, 10 and 14 (see legend at the bottom of the figure for color code) and different values of the threshold $\theta$: a,b) $\theta=0.2$, c,d) $\theta=0.6$. For comparison, in all panels we report also the curves of Fig.~\ref{fixed_vs_time_varying} (with dashed lines) and the global maximum of the curves of Fig.~\ref{fixed_damage} (with horizontal dotted lines).
    }
    \label{time_vs_mutating}
\end{figure}

The results are reported in Fig.~\ref{time_vs_mutating}, where $D/N$ is plotted vs. $\varepsilon$ for the case of a mutating virus.
One can see that the emerging qualitative scenario is similar to that of time varying viruses, in that one has an initial growth and an asymptotic behavior with a local maximum in between. Moreover, we again observe that for low thresholds a range of $\varepsilon $ values exists for which the injury to the system is higher than the case of a constant base damage whereas at high thresholds a virus with fixed base damage is more harmful.
The comparison with Fig.~\ref{fixed_vs_time_varying}, however, suggests that a virus with increasing base damage over time always produces a larger injury to the system than mutating virus.

In conclusion, we have introduced a novel compartmental model able to describe the spreading of a malware (and of the awareness of its incidence) on a given
network of devices. The novelty of our approach consists in having considered vectorial compartments made of two components,
the first being descriptive of the state of the device with respect to the virus propagation, and the second accounting for the awareness
of the device's user about the presence of the cyber threat. 
The model allows to follow the evolution of a cyber-epidemic from the moment at which a malware is seeded
in a network of devices, until when a given user gets aware of the incurred damage and starts a wave of
awareness which eventually leads to the development of a proper antivirus software.
We then illustrate the overall injury that a malware is able to produce in Erd\H{o}s-R\'enyi and scale-free architectures
for both the case in which the virus is causing a fixed damage on each device and the case in which, instead, the virus is
engineered to mutate while replicating from device to device.
As our world is more and more interconnected, a cyber-epidemic is a dangerous threat, where a cyber virus would undermine the global Internet system with catastrophic consequences. 
Our results are the attempt to describe, in a customized way, the evolution of such pandemic, and our conclusion may give hints on how to properly engineer a network of devices to minimize its vulnerability.

This work has been partially suppported by projects PGC2018-101625-B-I00 (Spanish Ministry, AEI/FEDER, UE) and M1993 (URJC Grant).
Authors acknowledge the usage of the resources, technical expertise and assistance provided by the supercomputing
facility CRESCO of ENEA in Portici (Italy).

%\begin{thebibliography}{00}%
%\end{thebibliography}

\bibliographystyle{unsrt}
% \bibliography{references}

\begin{thebibliography}{10}
\bibitem{specifications}
Our model has been simulated using {P}ython, and in particular the existing
  {N}etwork{X} library, with transition functions from one state to the other
  ruled by the respective probabilities. {U}nless otherwise specified, the
  reported numerical results refer to an ensemble average over 500 different
  network realizations of $n=1000$ nodes each, and a further ensemble average
  over 20 initial conditions per network (i.e., 20 different random positions
  where the initial seed of infected-unaware is placed).
\end{thebibliography}

\end{document}